\documentclass[preprint,12pt]{elsarticle}

\usepackage{graphics}
\usepackage{graphicx}

\usepackage{amssymb,amsmath}
\usepackage{amsthm}
\usepackage{bm}

\journal{Physics Letters B}

\begin{document}

\begin{frontmatter}



\title{Flavor neutrino oscillations in matter moving with acceleration}


\author[1,AIS]{Alexander I. Studenikin}
\ead{studenik@srd.sinp.msu.ru}
\author[1]{Ilya V. Tokarev}
\ead{tokarev.ilya.msu@gmail.com}

\address[1]{Department of Theoretical Physics, Faculty of Physics, Moscow State University, Moscow 119991, Russia}
\address[AIS]{Joint Institute for Nuclear Research, Dubna 141980, Moscow Region, Russia}

\begin{abstract}
A short review on studies of neutrino flavor oscillations in moving with a constant speed matter is presented.  Then neutrino flavor oscillations in matter moving with acceleration are considered. We develop an approach that is valid for both non-relativistic and relativistic matter motion. For both these cases the effective neutrino potentials in accelerating matter are obtained and   neutrino flavor oscillation probabilities and the Mikheyev-Smirnov-Wolfenstein resonance conditions are evaluated.  The obtained results in the limit of zero matter acceleration reproduce the corresponding expressions for the case of matter at rest. The developed approach is applied for neutrino flavor oscillations in the case of rotating matter. The results obtained might be of interest for astrophysical applications.
\end{abstract}


\end{frontmatter}


\section{Introduction}

It is well-known fact that astrophysical neutrino beams cover a great distances from sources of neutrinos through the space before reaching the earth's surface. But the most interesting changes of neutrino beams occurs during its propagation through the extreme background conditions, for instance, during propagation through the high density matter. This conditions are typical for astrophysical high intensive neutrino sources (solar neutrino, supernova neutrino and etc.). Thus studies of neutrino behavior and, in particular, studies of flavor neutrino oscillations in high density matter are of interest.\par
The theory of two-flavor neutrino oscillations in matter at rest has been developed by Wolfenstein \cite{Wolfenstein:1977ue}. Using the developed approach Mikheev and Smirnov \cite{Mikheev:1986gs} have obtained the most famous result in neutrino oscillation theory that predicts the resonance amplification of neutrino flavor oscillations in matter (the Mikheev-Smirnov-Wolfenstein effect)
\begin{equation}
\label{MSW_matter_at_rest}
\frac{\delta m^2}{2E}\cos2\theta=\sqrt{2}G_Fn_e.
\end{equation}
Here $n_e$ is density of electron matter component in the reference frame for which the total speed of matter is zero. Note that neutron and proton matter components due to weak interactions gives symmetric contributions to neutrino energies of different flavors. Thus only electron matter component effects on neutrino oscillations even in the most general case of matter structure (electrons, protons and neutrons). The obtained results are very crucial not only for the solar and atmospheric neutrino puzzles but also have important applications in different astrophysical and cosmology environments (see, for example, \cite{Raffelt:1996wa}).\par
The study of neutrino oscillations in the case of relativistic matter motion (for instance, relativistic jets of plasma from quasars) was done in \cite{Grigoriev:2002zr}. The resonance condition in the case of the relativistic unpolarized matter motion with constant velocity $\textbf{\textit{v}}$ has the form
\begin{equation}
\label{MSW_matter_const_vel}
\frac{\delta m^2}{2E}\cos2\theta=
\sqrt{2}G_F\frac{n_e}{\sqrt{1-\textbf{\textit{v}}^2}}(1-\boldsymbol{\beta}\textbf{\textit{v}}),
\end{equation}
where $\boldsymbol{\beta}$ is neutrino velocity. This result predicts increase of the oscillation probability for ultra-relativistic neutrinos in two cases: 1) the relativistic motion of matter directs along the neutrino propagation and the matter density $n_e$ is too high for the resonance appearance in non-moving matter; 2) the relativistic motion of matter directs in opposite direction to the neutrino propagation and the matter density $n_e$ is too low for the resonance appearance in non-moving matter.\par
The aim of our letter is the development of the theory of neutrino flavor oscillations in matter moving with acceleration. We consider two cases of matter motion both non-relativistic (including rotation) and relativistic. The most significant effects occur in the case of relativistic matter motion. Thus we hope that obtained results are of interest for understanding of neutrino beam evolution during, in particular, supernova explosion.\par
In next section we develop flavor neutrino oscillations theory in the case of relativistic matter moving with acceleration. Oscillation probability and the resonance condition are obtained. In the limit of matter at rest the results reproduce the solution obtained in \cite{Mikheev:1986gs} (see eq.(\ref{MSW_matter_at_rest})). The conditions of full transition of electron neutrinos to muon neutrinos are obtained. Particular, the angle $\phi_0$ between directions of matter and neutrino beam motion as a function of acceleration is determined. In the case of opposite directions of matter and neutrino beam motion ($\phi=\pi$) this condition fixes matter acceleration.\par
In conclusions using the developed approach we consider neutrino oscillations in the case of non-relativistic and rotating matter as another examples of accelerating matter. Some aspects and possible applications of obtained results are discussed.

\section{Flavor neutrino oscillations in matter moving with acceleration}

We consider neutrino propagating through the matter moving with acceleration. For simplicity, we develop the theory of two-flavor neutrino oscillations, e.g. $\nu_e\rightarrow\nu_{\mu}$, in matter moving by a constant force $\textbf{F}$ as a whole. Generalization for the case of other types of neutrino conversions could be done using the standard theory of three-neutrino mixing \cite{Giunti:2007ry}. It is assumed that neutrino start propagation through the matter simultaneously with the beginning of the acceleration.\par
Flavor neutrino interacts with matter through coherent forward elastic weak charge current and neutral current scatterings. The corresponding part of the neutrino effective Lagrangian describing flavor (electron or muon) neutrino interaction with matter could be written in the form
\begin{equation}
\label{lagrangian}
\mathcal{L}_{eff}=-\bar{\nu}_f\left(\gamma_{\mu}f^{\mu}\frac{1+\gamma^5}{2}\right)\nu_f.
\end{equation}
Matter parameter $f^{\mu}$ in the case of unpolarized matter is connected with the matter potential $V_f$
\begin{equation}
\label{matter_potential}
f^{\mu}=V_f(1,\textbf{\textit{v}}), \quad V_f=\frac{G_F}{\sqrt{2}}n_{\nu_f}.
\end{equation}
In the eqs.~(\ref{matter_potential}) $\textbf{\textit{v}}$ is matter velocity, and matter densities $n_{\nu_f}$ correspondingly for electron and muon neutrinos are given by
\begin{equation}
\label{densities}
n_{\nu_{e,\mu}}=n_e(4\sin^2\theta_W\pm1)+n_p(1-4\sin^2\theta_W)-n_n.
\end{equation}\par
The presence of matter modifies Dirac equation
\begin{equation}
\label{dirac_eq}
(\gamma_0E-\mathbf{\boldsymbol{\gamma}}\textbf{p}-m)\psi=
V_f(\gamma_0-\mathbf{\boldsymbol{\gamma}}\textbf{\textit{v}})\psi.
\end{equation}
The solution of the eq.~(\ref{dirac_eq}) introduces neutrino energy spectrum in matter
\begin{equation}
\label{energy_spectrum}
E=\sqrt{m^2+(\textbf{p}-\textbf{\textit{v}}V_f)^2}+V_f.
\end{equation}\par
Energy spectrum~(\ref{energy_spectrum}) in the case of relativistic neutrino can be represented as the sum of a free motion energy with potential energy $W_f$ arising from flavor neutrino interaction with moving matter
\begin{equation}
\label{potential_energy_full}
W_f=V_f\left(1-\frac{\textbf{p}\textbf{\textit{v}}}{\textrm{p}}+\frac{\textit{v}^2V_f}{2\textrm{p}}\right).
\end{equation}
Here we use notations $\textrm{p}=|\textbf{p}|$ and $\textit{v}=|\textbf{\textit{v}}|$. Note that the last term in the potential energy $W_f$ even in the case of high density relativistic matter gives a negligible contribution to the neutrino energy. Thus one can use indeed~(\ref{potential_energy_full}) the following expression
\begin{equation}
\label{potential_energy}
W_f\simeq V_f(1-\textit{v}\cos\phi),
\end{equation}
where $\phi$ is angle between neutrino and matter directions of propagations.\par
It is worse to stress that one should use the potential energy in full form~(\ref{potential_energy_full}) when $\phi=\frac{\pi}{2}$ (e.g., when $\textbf{p}\textbf{\textit{v}}=0$). This situation is considered it the end of our letter during discussion of neutrino propagation through the rotating matter.\par
The evolution equation of flavor states in the case of relativistic neutrinos can be written in the form (see also \cite{Pal:1991pm})
\begin{equation}
\label{evol_eq}
i\frac{d}{dx}\nu(x)=\left(
\frac{\Delta\sin2\theta}{2}\sigma_1-
\frac{\Delta\cos2\theta}{2}\sigma_3+W
\right)\nu(x), \quad
\nu(x)=\begin{pmatrix} \nu_e(x) \\ \nu_{\mu}(x) \end{pmatrix}.
\end{equation}
Here we use notation $\Delta=\delta m^2/2E$, where $\delta m^2=m_2^2-m_1^2$ is the difference of the neutrino masses squired ($m_{1,2}$ are masses of physical neutrino states), $E$ is neutrino energy, $\theta$ is vacuum mixing angle. Potential energy matrix $W$ is connected with flavor neutrino basis
\begin{equation}
\label{pot_en_matrix}
\nu_f=\delta_{fe}\begin{pmatrix} 1 \\ 0 \end{pmatrix}+
\delta_{f\mu}\begin{pmatrix} 0 \\ 1 \end{pmatrix}, \quad
W=\begin{pmatrix} W_e & 0 \\ 0 & W_{\mu} \end{pmatrix}.
\end{equation}\par
In the case of relativistic motion matrix (\ref{pot_en_matrix}) is given by
\begin{equation}
\label{pot_en_matrix_rel}
W=\frac{G_F}{\sqrt{2}}\frac{n_e}{\sqrt{1-\textit{v}^2}}(1-\textit{v}\cos\phi)\sigma_3,
\end{equation}
where electron density $n_e$ is given in the reference frame for which the total speed of
electrons is zero and $\gamma={(1-\textit{v}^2)}^{-\frac12}$ occurs to account for relativistic matter motion.\par
Let us underline that in eqs.~(\ref{evol_eq}) and (\ref{pot_en_matrix_rel}) terms proportional to unity matrix (including contributions from proton and neutron matter components) were neglected because they do not effect the oscillation probability.\par
Under a constant force $\textbf{F}$ matter moves as whole with velocity
\begin{equation}
\label{speed_rel}
\textit{v}(x)=\frac{ax}{\sqrt{1+(ax)^2}}, \quad a=\frac{\textrm{F}}{m_e},
\end{equation}
where $x$ is distance traveled.\par
Following the standard procedure for flavor neutrino oscillations we use eqs.~(\ref{pot_en_matrix_rel}) and (\ref{speed_rel}) to find the solution of the eq.~(\ref{evol_eq}), describing evolution of electron and muon flavor neutrinos
\begin{equation}
\label{evol_eq_norel}
\nu(x)=\exp\left\{-\frac{i}{2}\left[
\Delta\sin2\theta\sigma_1+
\left(\Delta\cos2\theta-A\right)\sigma_3
\right]x\right\},
\end{equation}
where theory parameter $A$ has the form
\begin{equation}
\label{A_r}
A=\frac{G_Fn_e}{\sqrt{2}}\left(\sqrt{1+(ax)^2}-ax\cos\phi+\frac{\ln\left(ax+\sqrt{1+(ax)^2}\right)}{ax}\right).
\end{equation}\par
The oscillations probability of $\nu_e\rightarrow\nu_{\mu}$ transition can be written in the form
\begin{equation}
\label{tran_prob}
P_{\nu_e\rightarrow\nu_{\mu}}(x)=|\nu^{\dag}_{\mu}\times\nu_e(x)|^2,
\end{equation}
and in the case of relativistic matter motion is given by
\begin{equation}
\label{tran_prob_r}
P_{\nu_e\rightarrow\nu_{\mu}}(x)=\sin^22\theta_{eff}\sin^2\frac{\pi x}{L_{eff}},
\end{equation}
where effective mixing angle $\theta_{eff}$ and effective oscillations length $L_{eff}$ are given by
\begin{equation}
\label{theta_r}
\sin^22\theta_{eff}=\frac{\Delta^2\sin^22\theta}
{\left(\Delta\cos2\theta-A\right)^2+\Delta^2\sin^22\theta},
\end{equation}
\begin{equation}
\label{L_r}
L_{eff}=\frac{2\pi}
{\sqrt{\left(\Delta\cos2\theta-A\right)^2+\Delta^2\sin^22\theta}},
\end{equation}\par
For certain matter density $n_e$ and neutrino energy $E$ we obtain the condition of the resonance amplification of neutrino flavor oscillations in relativistic matter moving with acceleration
\begin{equation}
\label{ampl_con_r_x}
\frac{\delta m^2}{2E}\cos2\theta=\frac{G_Fn_e}{\sqrt{2}}
\left(\sqrt{1+(ax)^2}-ax\cos\phi+\frac{\ln\left(ax+\sqrt{1+(ax)^2}\right)}{ax}\right).
\end{equation}
The condition~(\ref{ampl_con_r_x}) in the limit of zero matter acceleration reproduces the result obtained before ((\ref{MSW_matter_at_rest}), the MSW effect) but nevertheless has the distinguishing difference from the result above. Due to dynamic changes of matter motion the oscillations amplitude explicitly depends on neutrino traveled distance~$x$. The resonance condition can be represented in terms of matter velocity
\begin{equation}
\label{ampl_con_r_v}
\frac{\delta m^2}{2E}\cos2\theta=\frac{G_Fn_e}{\sqrt{2}}
\left(\frac{1-\textit{v}\cos\phi}{\sqrt{1-\textit{v}^2}}+
\frac{\sqrt{1-\textit{v}^2}}{\textit{v}}\ln\sqrt{\frac{1+\textit{v}}{1-\textit{v}}}\right), \quad
\textit{v}=\textit{v}(x).
\end{equation}\par
It should be stressed that matter density value which is not satisfied the resonance condition for matter at rest~(\ref{MSW_matter_at_rest}) can produce resonance amplification of oscillations due to matter accelerating motion. Further we shall discuss two important cases of relativistic matter motion.\par
In the case of matter motion along the direction of neutrino beam the theory parameter $A$ tends to zero, e.g. effect of matter vanishes. Thus if $\phi=0$ the probability~(\ref{tran_prob_r}) describes neutrino oscillations in vacuum
\begin{equation}
\label{res_con_zero_phi}
\theta_{eff}=\theta, \quad L_{eff}=\frac{2\pi}{\Delta}.
\end{equation}
This phenomenon was predicted for relativistic matter motion with constant velocity in~\cite{Grigoriev:2002zr} and also occurs in the case of relativistic matter motion with acceleration.\par
If $\phi\neq0$ the resonance condition~(\ref{ampl_con_r_x}) reduces for the following expression
\begin{equation}
\label{res_con_nonzero_phi}
\frac{\delta m^2}{2E}\cos2\theta=\sqrt{2}G_Fn_eax\sin^2\frac{\phi}{2}.
\end{equation}
Using the explicit dependence of the resonance condition~(\ref{res_con_nonzero_phi}) on the traveled distance $x$ one can determine the neutrino beam direction of full transition $\nu_e\rightarrow\nu_{\mu}$. This phenomenon can occur only for traveled distance $x_0=L_{eff}/2$ (see~(\ref{tran_prob})). The corresponding angle $\phi_0$ for certain values of matter acceleration and density we obtain in the form
\begin{equation}
\label{angle_full_tran}
\sin^2\frac{\phi_0}{2}=\frac{\Delta^2\cos2\theta}{\pi\sqrt{2}G_Fn_ea}, \quad
a>\left(\frac{1\,eV^2}{G_Fn_eE}\right)\frac{\textrm{m}}{\textrm{s}^2}.
\end{equation}
In particular, in the case of opposite direction of matter and neutrino beam motion ($\phi=\pi$) the expression~(\ref{angle_full_tran}) fixes matter acceleration
\begin{equation}
\label{accel_opp}
a_0=\left(\frac{1\,eV^2}{G_Fn_eE}\right)\frac{\textrm{m}}{\textrm{s}^2}.
\end{equation}\par
It is worse to stress that during neutrino propagation through the uniform layer of matter at the angle $\phi_0$ the point of full transition $x_0$ should be reached inside the matter. Thus, the thickness of accelerated matter layer should satisfy the condition: $h>x_0\tan\phi_0$.\par

\section{Conclusions}

In our letter we develop the theory of flavor neutrino oscillations in matter moving with acceleration. In the case of relativistic matter motion the conditions of the resonance amplification of neutrino flavor oscillations in matter for different neutrino beam directions are obtained. Particular, for $\phi=0$ the affect of matter vanishes, mixing angle and oscillations length correspond to the vacuum values. For $\phi\neq0$ the angle $\phi_0$ of full transition $\nu_e\rightarrow\nu_{\mu}$ is obtained. We predict the absence of electron neutrinos in neutrino beam after passing through the moving matter at the angle $\phi_0$.\par
The obtained results might be of interest in application to supernova evolution~\cite{2012A&ARv..20...49V}. Particular, during supernova core collapse a great amount of neutrinos are produced in central part of the star and escape through the core which collapses in on itself. In this case $\phi=\pi$ and condition~(\ref{accel_opp}) determines value of matter acceleration when full transition $\nu_e\rightarrow\nu_{\mu}$ can occur. Note, that shocks from supernova explosions propagating through the interstellar medium can account for the acceleration of galactic cosmic rays. Similar processes occurring in extragalactic radio sources can lead to acceleration of relativistic electrons~\cite{Blandford:1978ky}. In this case $\phi=0$ and matter influence becomes negligible (see (\ref{res_con_zero_phi})).\par
The developed theory could be applied to describe flavor neutrino oscillations in non-relativistic matter moving with acceleration and in rotating matter. We obtain the conditions of the resonance amplification of neutrino flavor oscillations in the following forms for non-relativistic matter
\begin{equation}
\label{A_nr}
A=\sqrt{2}G_Fn_e\left(1-\frac23\textit{v}(x)\cos\phi\right), \quad \textit{v}(x)=\sqrt{2ax},
\end{equation}
and for rotating matter
\begin{equation}
\label{A_rot}
A=\sqrt{2}G_Fn_e\left(1-\frac{G_F(n_n-n_e)}{3\sqrt{2}E}\textit{v}^2(r)\right), \quad \textit{v}(r)=\omega r.
\end{equation}
Note that in the case of rotating matter $\phi=\frac{\pi}{2}$ and one can use eq.~(\ref{potential_energy_full}) for potential energy. Also it is supposed that neutrino propagates only in the radial direction from the rotation axis.\par
Obviously, these two kinds of motion almost no effect on oscillations probability, and one can use results obtained for matter at rest. Only relativistic motion changes the features of the flavor neutrino oscillations in matter.

\section*{Acknowledgements}
This work was supported by the RFBR grant 11-02-01509.

\newpage

\bibliographystyle{unsrt}
\bibliography{flav_osc_in_acc_mat}






\end{document}